\documentstyle[aps,preprint]{revtex}
\renewcommand{\baselinestretch}{1.3}
\begin{document}
\title{Observability of relative phases of macroscopic quantum states}
\author{Arun Kumar Pati}
\address{Theoretical Physics Division, 5th Floor, Central Complex}
\address{Bhabha Atomic Research Centre, Mumbai-400 085, INDIA}
\date{\today}
\maketitle

\begin{abstract}
After   a   measurement,   to  observe  the  relative  phases  of
macroscopically distinguishable states  we  have  to  ``undo''  a
quantum  measurement.  We  derive  an  {\it  inequality} which is
satisfied   by   the   relative   phases    of    macroscopically
distinguishable  states  and  consequently  any  desired relative
phases can not be observed in interference setups. The  principle
of  {\it macroscopic complementarity} is invoked that might be at
ease with the macroscopic world. We illustrate the idea of  limit
on  phase  observability  in  Stern-Gerlach  measurements and the
implications are discussed.
\end{abstract}

\vskip3cm


\vskip 3cm

email:krsrini@magnum.barc.ernet.in

\newpage
\par
Although  the  principle  of linear   superposition  is  a  basic
property of microscopic quantum system it has been  debated  over
the  years that some macroscopic quantum system can also be found
to be in a superposition of  all  possible  states  (one  example
would  be  Schr\"odinger's Cat state). A Schr\"odinger Cat state is a
macroscopic object which may be in a {\it linear superposition of
states} corresponding to macroscopically different beings (living
and dead) \cite{1}. However, when an observation is made  on  the
Cat  state  it  is found either living or dead but not both. Thus
extending the  linearity  of  quantum  mechanics  to  macroscopic
domain  conflicts  with  macroscopic  realism. But one can always
question the validity of the linear  superposition  principle  of
quantum  mechanics  when  it  is  applied to complicated physical
systems consisting of large  number  of  atoms  and  molecules  or
physical  systems  of  macroscopic dimension. For example Leggett
\cite{2} has discussed this question in detail and other  issues
such  as  experimental  support  in favour of macroscopic quantum
phenomena. However, we do not address such issues  here.  Rather,
we  assume  that  in  nature  physical  systems do exist in linear
superposition of macroscopically distinguishable (MD) states.
Some examples of this are superconducting quantum interference devices
\cite{3}, the possibilty of optical production of Cat states \cite{4} and  recently
discovered  Bose-Einstein condensates \cite{5}. Infact, the issue
of detection of relative phase \cite{6} between two BE condensate
and coherent quantum tunneling between two BE condensates \cite{7}
have been discussed in literature.  If  the  macroscopic  quantum
system  is  in  a  superposed  state, then the phase relationship
between macroscopically distinguishable states must  be  observed
through  interference.  Hence  it is of fundamental importance to
discuss the issue of observability of the relative phases of  the
macroscopically  distinguishable  states  and limit for realising
such superposition states.

\par
Any effort to observe interference effect for macroscopic systems
would  be  a  difficult  task experimentally, because macroscopic
systems often interact dissipatively in a irreversible  way  with
the  environment  \cite{8}  and  this  causes  the  loss of phase
coherence  between  different  branches  of  macroscopic  states.
Another  argument  against  observing relative phases is that the
measuring apparatus is so large in size that it is impossible  to
distinguish  between  the  pure state of the total system (system
plus apparatus)  and  the  statistical  mixture  \cite{9}.  Peres
\cite{10}   has   argued   that   observing  relative  phases  of
macroscopically distinguishable states  requires  measurement  of
certain  explicit  time-dependent  operators  and  measurement  of
classical  analog  of  the  operator  would   violate   classical
irreversibility.  It is therefore apparent that all the arguments
against  observing  phases   of   macroscopic   states   involve
irreversibility  in  some  form  or  the other and is stated as a
sufficient condition for the  loss  of  phase  coherence  between
different branches of the macroscopic states. However, there  has
not  been any attempt to answer the question of (un)observability
of the relative phases of macroscopic systems  using  fundamental
principles of quantum theory.

\par
In this paper we raise an important  and fundamental question: Is
there an intrinsic quantum limit  on  the  observability  of  the
phases  of  macroscopically  distinguishable states? It turns out
that there is a {\it limit on the observability of the phases  of
macroscopic states}. The inequality that we derive would put some
restriction  on the phase relationship between different branches
of   macroscopically   distinguishable   states   and   on    the
observability  of  the  off-diagonal  matrix elements of the pure
state density operator. Since the  presence  of  phase  coherence
distinguishes a pure state from a mixture {\it the new inequality
on  phase  relationship will be a criterion for the purity of the
macroscopic quantum states}. Further, we invoke a {\it  principle
of  macroscopic complementarity} which would lead to emergence of
classical reality for sufficiently large macroscopic systems.  We
apply  these ideas for Stern-Gerlach type measurement and discuss
the implication of the inequality for realising the superposition
of macroscopically distinguishable states.

\par
In the forgoing  discussion  unlike Bohr's doctrine           the
measurement apparatus is  not  treated  classically  and  quantum
mechanical  laws  are applied to the system as well as apparatus.
Suppose we design  an  apparatus  to  measure  the  value  of  an
observable  $O$  of a system. The state of the system $|\phi> \in
{\cal H}_s$ has its own eigenstates $\{|\phi_n>\}$ which forms  a
complete  set  and eigenvalues $\{O_n\}$ and the initial state of
the apparatus is $|\psi> \in {\cal H}_a$ . If the  initial  state
of  the system is in a linear superposition of the form $|\phi> =
\sum_n c_n|\phi_n>$, then the combined state of the  system  plus
apparatus can be written as $\sum_n c_n|\phi_n>\otimes|\psi>$. As
a  result  of  interaction  between  the system and apparatus the
final state of the apparatus is different from $|\psi>$  and  the
combined state is written as $\sum_n c_n|\phi_n>\otimes|\psi_n>$.
If the apparatus pointer has to associate a distinct and definite
eigenvalue  $O_n$  of  the  observable  of  the  system  to  some
apparatus state $|\psi_n>$, then on physical  ground  we  require
the  states  $\{|\psi_n>\}$  should  be  mutually  orthogonal and
macroscopically distinguishable \cite{11}. Thus  the  macroscopic
apparatus  during  the  interaction  process  has  evolved into a
linear superposition of macroscopically  distinguishable  states.
Now,    the    observation    of   phase   relationship   between
macroscopically  distinguishable  states  would  mean   observing
quantities  like  $c_i^*c_j  (i \ne j)$ which are nothing but the
off-diagonal matrix elements of the pure state  density  operator
$\rho$, i.e. $(\rho_{ij}) = c_i^*c_j$, since interference effects
are  contained  in  the  off-diagonal  elements  of  the  density
operator.

\par
Let  us consider a  general model of measurement where we want to
measure an observable $O$ of a system attached to  a  macroscopic
apparatus.  The  pointer  of  the  apparatus has (center-of-mass)
coordinate $q$, conjugate momentum $p$  and  initially  localised
arround  $q  =  0$.  The pointer has to move a macroscopic length
$L_n$ to measure the eigenvalue $O_n$, where $L_n  =  LO_n$.  The
total   Hamiltonian   of  the  combined  system  would  be  $H_T  =
H_s \otimes 1_a  +  1_s \otimes H_a + H_c$, where $H_s$ and $H_a$ are
system and apparatus Hamiltonian, respectively. During the interaction only
the coupling Hamiltonian  $H_c  =  V(t)Op$  is  important,  where
$V(t)$  is very large quantity (velocity of the pointer) and $L =
\int V(t) dt$. We can write the combined state of the system  and
apparatus        before        interaction       as       $\sum_n
c_n|\phi_n>\otimes|\psi>$. In position  representation  the  wave
function   of  the  apparatus  could  be  $<q,q_2,...q_N|\psi>  =
\psi(q,q_2,...q_N)$ where  $q_2,...q_N$  are  regarded  as  large
number of ``irrelevant''   or   ``inactive''  degrees  of  freedom
\cite{10}. After coupling the state of the combined system is
\begin{equation}
|\Psi> = \sum_n c_n |\phi_n>\otimes e^{-iLpO_n} |\psi>.
\end{equation}

\par
To  measure  the eigenvalue $O_n$ we have to look for the pointer
position coordinate $q$. The expectation value of the  observable
$O$   is  still  given  by
\begin{equation}
<\Psi|O|\Psi> = \sum_n|c_n|^2 O_n.
\end{equation}
The  pointer  of  the  apparatus  would  be  at  $q  =  L_n$ with
probability  $|c_n|^2$.  Now   we   discuss   the   question   of
observability   of   relative  phase  of  different  branches  of
macroscopically distinguishable states after  the  interaction of
the system and apparatus is over. We look for the operator  whose
expectation  value  would give the information about the relative
phases of different branches.

\par
To derive the limit on the relative  phases of different branches
we introduce two hermitian operators which is a  measure  of  the
phases   of   the   ith   and   jth   branches.   Let  us  define
$A_1$ and $A_2$ as
\begin{eqnarray}
A_1 =
{1\over2}(e^{-iL_ip/\hbar}|\phi_i><\phi_j|e^{iL_jp/\hbar}       +
e^{-iL_jp/\hbar}|\phi_j><\phi_i|e^{iL_ip/\hbar})        \nonumber
\end{eqnarray}

\begin{equation}
A_2 =  {i\over2}(e^{-iL_jp/\hbar}|\phi_j><\phi_i|e^{iL_ip/\hbar}
-  e^{-iL_ip/\hbar}|\phi_i><\phi_j|e^{iL_jp/\hbar}),
\end{equation}
with $A_1^2 = A_2^2 = {1 \over 4}(P_i +  P_j)$  where  $P_i$  and
$P_j$  are  the projection operators corresponding to eigenstates
$|\phi_i>$  and  $|\phi_j>$.  The  expectation  value  of   these
operators  in  the  state  $|\Psi>$ just after the measurement is
given by
\begin{eqnarray}
<\Psi|A_1|\Psi>   =     {1\over2}(c_i^*c_j    +    c_ic_j^*)    =
|c_i|c_j|\cos\phi_{ij}  \nonumber
\end{eqnarray}

\begin{equation}
<\Psi|A_2|\Psi> =  {i\over2}(c_ic_j^* - c_i^*c_j) = |c_i|c_j|\sin\phi_{ij},
\end{equation}
where  $\phi_{ij}$  are  the  relative  phases  of  ith  and  jth
branches.

\par
To observe the interference pattern we have to wait for some time
after  the  system  has  interacted  with  the apparatus and then
superpose its different branches. As a result the combined  state
is  no longer described by $|\Psi>$. At later time the state will
evolve under the action of the Hamiltonian $H = H_s \otimes 1_a +
1_s \otimes H_a$. We assume that the observable $O$ of the system
commutes with the system Hamiltonian $H_s$. The evolved state  is
now given by

\begin{equation}
|\Psi(t)> = e^{-{itH/\hbar}} |\Psi> =  \sum_n c_n |\phi_n>\otimes e^{-itH/\hbar} e^{-iL_np} |\psi>.
\end{equation}
The   expecation   value   of   the   operators $A_1$ and $A_2$ in  the  state
$|\Psi(t)>$ are given by
\begin{eqnarray}
<\Psi(t)|A_1|\Psi(t)> =  {1\over2}(c_jc_i^*  <\psi|e^{iL_ip}  e^{itH}  e^{-iL_ip}
e^{iL_jp} e^{-itH} e^{-iL_jp}|\psi> + c.c.) \nonumber
\end{eqnarray}

\begin{equation}
<\Psi(t)|A_2|\Psi(t)> =  {i\over2}(c_ic_j^* <\psi|e^{iL_jp}  e^{itH}  e^{-iL_jp}
e^{iL_ip} e^{-itH} e^{-iL_ip}|\psi> - c.c.).
\end{equation}
This can be written as
\begin{equation}
<A_1> = |c_i||c_j||Z_{ij}(t)| \cos\Phi_{ij}(t)~~~~~~
<A_2> = |c_i||c_j||Z_{ij}(t)| \sin\Phi_{ij}(t),
\end{equation}
where $Z_{ij}(t) = <\psi|e^{itH(q+L_i)/\hbar}
e^{-itH(q+L_j)/\hbar}
|\psi>$ and
$e^{iL_ip/\hbar} e^{itH/\hbar} e^{-iL_ip/\hbar} =  e^{itH(q+L_i)/\hbar}$
etc.   Here,  $H(q+L_i)$  and  $H(q+L_j)$  are  nothing  but  the
Hamiltonian $H$ with coordinate $q$ has been shifted by an amount
$L_i$ and $L_j$. The phase $\Phi_{ij}(t)$ is relative  phases  of
the  different branches of macroscopically distinguishable states
at any time t which contains the phases of  $c_i^*c_j$
as well as that of $Z_{ij}(t)$. Note  that  $Z_{ij}(t)$  contains
macroscopic  parameters  such  as  length  $L$,  mass  $M$ of the
pointer apparatus.

To  what  extent the relative phase information we can retrieve is
given by the limit that we derive below on the relative phase  of
two  branches  of  the macroscopically distinguishable states. We
will  show that it is not possible to observe any arbitrary phase
relationship. Macroscopic states only with certain relative phase
difference can be observed in interference set ups. We apply  the
generalised  uncertainty  relation to two non-commuting hermitian
operators $A_1$ and $A_2$. This is given by

\begin{equation}
{\Delta  A_1}^2  {\Delta  A_2}^2  \ge  {1\over4}|<[A_1, A_2]>|^2.
\end{equation}

\par
We evaluate the uncertainties in the state $|\Psi(t)>$ of     the
combined system at an arbitrary time t. They are given by

\begin{eqnarray}
{\Delta A_1}^2 = <A_1^2> - <A_1>^2 = {1\over4}(|c_i|^2 + |c_j|^2) - |c_i|^2|c_j|^2|Z_{ij}(t)|^2 \cos^2\Phi_{ij}(t) \nonumber
\end{eqnarray}
\begin{equation}
{\Delta A_2}^2 = <A_2^2> - <A_2>^2 = {1\over4}(|c_i|^2 + |c_j|^2) - |c_i|^2|c_j|^2|Z_{ij}(t)|^2 \cos^2\Phi_{ij}(t)
\end{equation}
and the expectation value of the commutator is given by
\begin{equation}
<\Psi(t)|[A_1,A_2]|\Psi(t)> = {i\over2}(|c_j|^2 - |c_i|^2)
\end{equation}
With  the  help  of  (9) and (10) we  can  simplify the inequality for the
relative phases as
\begin{equation}
\sin^22\Phi_{ij}(t) \ge {(|c_i|^2 + |c_j|^2)|Z_{ij}(t)|^2 - 1 \over |c_i|^2|c_j|^2|Z_{ij}(t)|^4}
\end{equation}
which gives the desired limit on the relative phases of different
branches  of  macroscopically distinguishable states.  In  actual
interferometry one generally measures the relative phase shift as
a  function  of   $\sin\Phi_{ij}$   and   not   $\sin2\Phi_{ij}$.
Therefore, we derive an inequality which expresses this fact.
The inequality that we derive below is a stronger one.
To further tighten the inequality for relative phases we make use of the
so called ``triangle  inequality''.  We  know  that  given  three
arbitrary   non-otrhogonal   vectors   $|\Psi_1>,  |\Psi_2>$  and
$|\Psi_3>$ belonging to the Hilbert space ${\cal H} = {\cal H}_s
\otimes {\cal H}_a$ we have  the
following triangle inequality

\begin{equation}
D(\Psi_1, \Psi_2) +  D(\Psi_2, \Psi_3) \ge  D(\Psi_1, \Psi_3),
\end{equation}
where  $D(\Psi_{\mu},  \Psi_{\nu}),  (\mu,\nu  =  1,2,3)$  is the
metric  defined  from  the  inner  product  between  the  vectors
$|\Psi_{\mu}>$  and  $|\Psi_{\nu}>$.  The  metric is a measure of
distance  \cite{12}  between  the  vectors   $|\Psi_{\mu}>$   and
$|\Psi_{\nu}>$ defined on the projective Hilbert space ${\cal P}
= {\cal H} /U(1)$ of combined system. For non-normaliseable
vectors we define it as \cite{13}

\begin{equation}
D(\Psi_{\mu}, \Psi_{\nu}) = \Biggl(1 -   {|<\Psi_{\mu}|\Psi_{\nu})>|^2 \over ||\Psi_{\mu}||^2||\Psi_{\nu}||^2}\Biggr).
\end{equation}
where  $||\Psi_{\mu}||^2$   is   the   norm   of    the    vector
$|\Psi_{\mu}>$  and
similarly for $|\Psi_{\nu}>$. This is   the  Fubini-Study  metric
which   is   invariant   under   all   unitary  and  anti-unitary
transformations acting on the Hilbert space ${\cal H}$. Now define three vectors as follows:
$|\Psi_1> = |\Psi(t)>, |\Psi_2> = A_1|\Psi(t)>$ and $|\Psi_3> = A_2|\Psi(t)>$.
The distance functions are given by

\begin{eqnarray}
D(\Psi_1, \Psi_2) = 1 - {4|c_i|^2|c_j|^2|Z(t)|^2  \cos^2\Phi_{ij}
\over (|c_i|^2 + |c_j|^2)}~~~~~  D(\Psi_2, \Psi_3) = 1
- {(|c_i|^2 - |c_j|^2)^2 \over (|c_i|^2 + |c_j|^2)^2},    \nonumber
\end{eqnarray}

and
\begin{equation}
D(\Psi_1, \Psi_3) = 1 - {4|c_i|^2|c_j|^2|Z(t)|^2\sin^2\Phi_{ij} \over (|c_i|^2 + |c_j|^2)}.
\end{equation}
Inserting (14) in (12) and simplifying the inequality we have
\begin{equation}
\cos2\Phi_{ij} \le {1 \over |Z(t)|^2(|c_i|^2 + |c_j|^2)}
\end{equation}
Further, on combining (15) with (11) we have the tightened version of  the
inequality  for  the  relative  phases  of the two branches as is
given by
\begin{equation}
(\sin^2\Phi_{ij})  \ge  {1 \over 2}\Biggl({|Z_{ij}(t)|^2(|c_i|^2 + |c_j|^2)   -  1  \over   |Z_{ij}(t)|^2(|c_i|^2 + |c_j|^2)    +
1}\Biggr) {(|c_i|^2 + |c_j|^2) \over |c_i|^2|c_j|^2|Z_{ij}(t)|^2}
\end{equation}

\par
The  above  inequality is more stronger than (11) as it makes use
of uncertainty relation and  triangle  inequality.  To  know  the
bound on the phases we really do not have to measure any operator
for that matter. The knowledge of probability distributions (i.e.
$|c_i|^2$  and  $|c_j|^2$)  and  expectaion  value  of  displaced
unitary operators (it contains the macroscopic parameters such as
length, mass and possibly other  variables  of  the  pointer)  is
enough to tell us the phase information. This inequality is valid
for  all  times even much after the measurement of the observable
$O$  of  the  system.  It  restricts  the   relative   phase   of
macroscopically distinguishable states and {\it those macroscopic
states  with  a relative phase lower than the predicted value can
not be superposed to produce an  interference  pattern}.  Because
then  that  would  reveal  the  observable  phase  information in
contradiction with the uncertainty (inequality)  principle.  This
is   a   {\it necessary} criterion  for  any  macroscopically
distinguishable states for producing an interference pattern.  If
the  relative  phases  of  different  branches  violate the above
inequality then the quantum interference effect is not  important
for all practical purposes.

\par
We can understand  the  limitation  if we renunciate the extended
Bohr's complimentarity for macroscopic systems  as  follws:  {\it
The  non-violation  of  the  inequality  by  relative  phases and
acquisition  of  which-state  information  (assiging  a  definite
macrostate)   are  mutually  exclusive.}  This  we  call  as  new
macro-complimentarity. Jammer \cite{14} has analysed Bohr's  view
on  macro-complimentarity  and  concluded  that the issue between
realism and idealism is  a  matter  subject  to  complimentarity.
Leggett  \cite{15} has discussed the possibility of exhibiting no
interference  between  macroscopically  distinct  states  and  in
assigning a definite state to a macroscopic system. The principle
of macroscopic complementarity consolidates the spirit of earlier
works in these lines.

\par
Does the macro-complimentarity shed some light on the macroscopic
systems  which are under every day-level observation. For example
why do we see a real Cat in one or the another state and not in a
superposition of different possible states.  It  seems  that  for
sufficiently  large  bodies  {\it the relative phases of distinct
branches are adjusted so as to violate the strong inequality (16)
thereby not allowing to observe the interference  between  them}.
Also,  it  can be verified that if we can assign a definite state
to a macrosystem the inequalities (11)  and  (16)  are  violated.
Thus,  the  new  inequality  together  with macro-complimentarity
helps us to understand the macroscopic world as we live. It might
be true that a real Cat is in a linear superposition  of  all  of
its  possible  states (living and dead) but the relative phase of
the  two  branches  would  not  satisfy  either  the  uncertainty
inequality  or  the  strong  inequality.  Thus,  the principle of
macroscopic complementarity helps us to understand the  emergence
of realism from the quantum mechanical principles.

\par
If  we  observe  the  relative  phase after a short time t of the
interaction then  we  may  assume  the  relative  phases  between
different   branches   may   be   small   and   one   can   write
$\sin\Phi_{ij}(t)$  as  $\Phi_{ij}(t)$.  Then  from   the   above
inequality  one can infer a minimum relative phase as is given by
\begin{equation}
(\Phi_{ij}(t)^2)_{min}     =     {1     \over
2}\Biggl({|Z_{ij}(t)|^2(|c_i|^2    +    |c_j|^2)    -   1   \over
|Z_{ij}(t)|^2(|c_i|^2 + |c_j|^2) + 1}\Biggr) {(|c_i|^2 + |c_j|^2)
\over |c_i|^2|c_j|^2|Z_{ij}(t)|^2}
\end{equation}

Is  there  any  limit  on  the  relative  phases  just  after the
interaction with the apparatus.  From  (16)  we  can  see  that
immediately after interaction the state of the combined system is
$|\Psi>$.  Hence,  the uncertainty inequality has to be evaluated
in the state $|\Psi>$ just after the interaction. The  inequality
is  given  by
\begin{equation}
\sin^2\phi_{ij} \ge {1 \over 2} \Biggl({(|c_i|^2 +
|c_j|^2)   -   1   \over   (|c_i|^2 + |c_j|^2) + 1}\Biggr) {(|c_i|^2 + |c_j|^2).
\over |c_i|^2|c_j|^2|Z_{ij}(t)|^2}
\end{equation}
The   significance   of   the  above  inequality  is  that  even
immediately after preparing a  superposition  of  macroscopically
different  states  we cannot obtain the phase information in any
desired way. There is an intrisic  quantum  limitation  in  doing
that.

\par
We  illustrate  the idea of limitation on the phase with the help
of  a  better  known  example-the Stern-Gerlach like measurement.
This example has been considered by Peres \cite{10} in trying  to
understand  the  connection  between  the  quantum mechanical and
classical irreversibilty. Consider a macroscopic apparatus  which
is  designed to measure the z-component of the spin-${1 \over 2}$
particle (say an electron). The  pointer  of  the  apparatus  has
(center-of-mass)  coordinate  $q$,  conjugate  momentum  $p$  and
initially localised arround $q = 0$. The motion of the pointer to
right or left will decide whether the spin is ${1  \over  2}$  or
$-{1   \over   2}$.   The   interaction  Hamiltonian  is  $H_c  =
V(t)\sigma_zp  =  2V(t)s_zp$  where  $s_z  =   {1\over2}\sigma_z,
\sigma_z$  being the Pauli spin matrix, $V(t)$ is velocity of the
pointer and $L = \int V(t) dt$ (a macroscopic distance).  We  can
write  the  initial  state of the spin-half particle as $|\phi> =
(\alpha|+> + \beta|->)$  and  of  apparatus  as  $|\psi>$.  After
coulping the state of the combined system is

\begin{equation}
|\Psi> = \alpha|+>\otimes e^{-iLp/\hbar} |\psi> +  \beta|->\otimes e^{iLp/\hbar} |\psi>
\end{equation}

\par
To  measure  the  z-component of the spin we have to look for the
sign of the pointer  position  coordinate  $q$.  The  expectation
value  of the $s_z$ is given by \begin{equation} <s_z> = {1 \over
2}(|\alpha|^2 - |\beta|^2)  \end{equation}  The  pointer  of  the
apparatus  would  be at $q = L$ with probability $|\alpha|^2$ and
at $q = -L$ with probability $|\beta|^2$. Once a  measurement  is
done  we  know only of $|\alpha|^2$ and $|\beta|^2$.
Here, we discuss the limit on the  relative  phases  of  the  two
branches from a fundamental uncertainty principle.

\par
As discussed earlier at  later  time  the state will evolve under
the action of the Hamiltonian $H = H_e + H_a$, where $H_e$ is the
elcetron Hamiltonian and $H_a$ is the apparatus Hamiltonian.  The
evolved state is now given by
\begin{equation}
|\Psi(t)> = e^{-itH/\hbar} |\Psi> =  \alpha|+>\otimes e^{-itH/\hbar} e^{-iLp/\hbar} |\psi> +  \beta|->\otimes e^{-itH/\hbar} e^{iLp/\hbar} |\psi>
\end{equation}
The operators $A_1$ and $A_2$ takes the simple form
\begin{equation}
A_1 = s_x \cos2Lp + s_y \sin2Lp ~~~~
A_2  =  s_x  \sin2Lp  - s_y \cos2Lp
\end{equation}

\par
We evaluate the uncertainties in the state $|\Psi(t)>$ of     the
combined system at an arbitrary time t. They are given by
\begin{eqnarray}
{\Delta A_1}^2 = <A_1^2> - <A_1>^2 = {1\over4} - |\alpha|^2|\beta|^2|Z(t)|^2 \cos^2\Phi \nonumber
\end{eqnarray}
\begin{equation}
{\Delta A_2}^2 = <A_2^2> - <A_2>^2 = {1\over4} - |\alpha|^2|\beta|^2|Z(t)|^2 \sin^2\Phi.
\end{equation}
where $|Z(t)|$ is given by  $< e^{itH(q-L)}  e^{-itH(q+L)}> =
|Z(t)|e^{i\theta(t)}$  and $\Phi(t) = \phi + \theta(t)$, $\phi$ being the relative phase
of $\alpha$ and $\beta$. Therefore the uncertainty inequality can be expressed as
\begin{equation}
\sin^22\Phi \ge {|Z(t)|^2  - 1 \over |\alpha|^2|\beta|^2|Z(t)|^4}
\end{equation}

\par
We   can express the tightened version of the inequality (16) for
the  relative  phases  of the two branches in Stern-Gerlach measurement as
\begin{equation}
(\sin^2\Phi)  \ge {1 \over 2} \Biggl({|Z(t)|^2   -  1  \over   |Z(t)|^2    +
1}\Biggr) {1 \over |\alpha|^2|\beta|^2|Z(t)|^2}
\end{equation}
which gives the desired lower bound on the relative phases of two
macroscopically  distinguishable  states.

\par
Some   consequence  of  the  above  inequality  relation  can  be
discussed now. If we ask what is the detectable phase just  after
the  ``premeasurement''.  In  that  case the inequality has to be
evaluated with the state (18) and $Z(t)$ is just  equal  to  one.
The inequality says that $\sin^2\phi \ge 0$ which is trivial. The
same  would  be true if the apparatus Hamiltonian does not depend
on $q$. Therefore for a non-trivial lower bound on  the  relative
phase we require the state to evolve for an appreciable period of
time and it is necessary that the Hamiltonian $H$ depends on $q$.
The  dependence  of $H$ on $q$ means the pointer of the apparatus
is moving under some  potential  $V_a(q)$  and  the  energy  thus
varies from place to place as time progresses. Interestingly, one
can  check  that if we assign a definite state to the macroscopic
system (say $|\alpha|^2 = 1$ and $|\beta|^2 = 0$) the  inequality
(25)   is   violated.  This  is  in  agreement  with  macroscopic
complementarity, stated earlier.

\par
Here,  we  briefly  discuss  the  argument of Peres for undoing a
quantum measurement and show that it is not a  serious  objection
against  observing relative phases. His argument runs as follows:
Let us define an operator $A = A_1 + i A_2$  and  the  expecation
value of the operator $A$ in this state is given by

\begin{equation}
<A> = \alpha \beta^* \int \psi^* e^{-iLp/\hbar} e^{itH/\hbar} e^{2iLp/\hbar} e^{-itH/\hbar} e^{-iLp/\hbar} \psi d^Nq
\end{equation}
This can be written as
\begin{equation}
<A> = \alpha \beta^* < e^{itH(q-L)/\hbar}  e^{-itH(q+L)/\hbar}>
\end{equation}
It  can be shown that for short time (for a proper choice of time
), $<A>$ can go to zero  and  the  relative  phases  of  the  two
macroscopic  states  is lost after some finite time. However, one
can measure another operator $A'$, where

\begin{equation}
A' = e^{-itH/\hbar} A e^{itH/\hbar}
\end{equation}
and  its  expectation  value  is nothing but $\alpha \beta^*$. But
such  an  operator  is  explicitly  time  dependent  constant  of
motion.  Classically (for a system with $N$ degrees of freedom we
have $2N$ constants of motion)) such  constants  of  motions  are
large  compared  to  constants  of  motion  which  are explicitly
time-independent. Such constants of motion are of no  use  to  us
because they are quite complicated for large $N$ and finite time.
This  results  in  unpredictability  of  the initial position and
momentum and hence in irreversibility. Therefore, Peres concludes
that the measurement of classical analog of operator (which gives
the relative  phases)  would  mean  the  violation  of  classical
irreversibility.

\par
But I believe that such argument against  ``undoing''  a  quantum
measurement is not a serious one. First of all not every  quantum
mechanical  operator has a classical analog although the converse
is true. The best example is the spin of a particle which has  no
classical analog. Indeed the operator one would measure to reveal
the  relative  phase  is  related  to  the components of the spin
operator and  it  is  not  expected  to  have  classical  analog.
Therefore  any  violation  in  classical world would not prohibit
``undoing'' a quantum measurement. Further, it has been proved by
Wigner \cite{16} that an operator which does not commute  with  a
conserved  quantity  can not be measured exactly (in the sense of
von Neumann).  In  the  discussion  of  Peres  {\it  neither  the
operator  $A$  nor $A'$ commute with a conserved quantity $s_z$}.
The z-component of the electron is a conserved quantity  assuming
that  $H_e$ does not contain any spin copmpnent other than $s_z$.
Therefore, even in principle the operator $A$ or $A'$ can not  be
measured exactly. As a result the exact relative phase can not be
obtained  by measuring the explicit time-dependent operator $A'$.
Hence, one should look for an estimate of the relative  phase  of
the  two  branches, which is precisely what we have aimed at.

\par
Thus,  in  conclusion  we have   discussed  the  limitations  for
realising macroscopic quantum superpositions. We have derived  an
{\it  inequality}  concerning  the  observability of the relative
phase of macroscopically distinguishable states. {\it  If  linear
superposition  principle  holds  for  macroscopic states then the
inequality has to  be  necessarily  satisfied}  by  the  relative
phases. It is suggested that the new inequality can be taken as a
criterion  for  the  purity  of  a  macroscopic quantum state. We
invoked the idea  of  macro-complimentarity  which  may  help  to
understand  how  does  a  macroscopic system come into a definite
state  and  it  may  resolve  the  issue  of  unobservability  of
interference  between  different  possible  (real) Cat states. We
argued that violation of classical irreversibility is not  always
(at  leat  in  the  example considered) serious objection against
``undoing'' a quantum measurement.

~~~~~~~~~~~~~~~~\\
Acknowledgements: I thank Dr. S. V. Lawande, Dr.C. S. Unikrishnan
and Prof. H. Rauch  for useful discussions.

\noindent
\newpage
\renewcommand{\baselinestretch}{1.3}

\end{document}